# Deeper Chandra Follow-up of Cygnus TeV Source Perpetuates Mystery

Yousaf Butt[1], Jeremy Drake[1], Paula Benaglia[2], Jorge Combi[3], Thomas Dame[1], Francesco Miniati[4], Gustavo Romero[2]

[1] *Harvard-Smithsonian Center for Astrophysics, 60 Garden St., Cambridge, MA 02138*
[2] *Instituto Argentino de Radioastronomía, CC5, (1894) Villa Elisa, Buenos Aires, Argentina and Facultad de Ciencias Astronómicas y Geofísicas, Universidad Nacional de La Plata, Paseo del Bosque, 1900 La Plata, Argentina*
[3] *Departamento de Física (EPS), Universidad de Jaén, Campus Las Lagunillas s/n, 23071 Jaén, Spain*
[4] *Physics Department, ETH Zürich, CH-8093 Zurich, Switzerland*

*revised version* 3 November 2005

## ABSTRACT

A 50 ksec Chandra observation of the unidentified TeV source in Cygnus reported by the HEGRA collaboration reveals no obvious diffuse X-ray counterpart. However, 240 Point-like X-ray sources are detected within or nearby the extended TeV J2032+4130 source region, of which *at least* 36 are massive stars and 2 may be radio emitters. That the HEGRA source is a composite, having as counterpart the multiple point-like X-ray sources we observe, cannot be ruled out. Indeed, the distribution of point-like X-ray sources appears non-uniform and concentrated broadly within the extent of the TeV source region. We offer a hypothesis for the origin of the very high energy gamma-ray emission in Cyg OB2 based on the local acceleration of TeV range cosmic rays and the differential distribution of OB *vs.* less massive stars in this association.

## 1. Introduction

The High Energy Gamma-Ray Astronomy (HEGRA) collaboration originally reported an apparently steady and extended TeV range gamma-ray source (TeV J2032+4130) in the Cygnus region based on about 121 hours of data collected between 1999 and 2001 using their stereoscopic Cherenkov telescopes (Aharonian et al., 2002). This source had no known lower-frequency counterparts. A further ~158 hours of observation from 2002 confirmed the existence of this mysterious source as well as its extended and steady nature (Aharonian et al., 2005). A reanalysis of data taken between 1989 and 1990 by the Whipple imaging Cherenkov telescope collaboration indicates that this source may, however, be variable on multi-year timescales – although, given the large uncertainties (of order ~30%), the older Whipple flux (0.12 Crab) and the recent HEGRA flux (0.05 Crab) are not inconsistent (Lang et al., 2004).

In an attempt to better understand this source, we had earlier carried out short Chandra[1] (5 ksec) and VLA[2] (8 minutes) observations but were unable to identify any firm counterparts (Butt et al. 2003). In that paper we proposed some possible explanations of this mysterious object based on its location within the massive Cygnus OB2 stellar association and concluded that the weak X-ray and radio emission from the TeV source region favored a nucleonic rather than electronic origin of the very high energy gamma-ray flux.

---

[1] For an overview of the Chandra X-ray Observatory please see eg., Weisskopf (2003) and also http://cxc.harvard.edu/
[2] The VLA is operated by the National Radio Astronomy Observatory (NRAO), which is a facility of the National Science Foundation (NSF), operated under cooperative agreement by Associated Universities, Inc.

Here we report on a deeper (50ksec) Chandra follow-up observation of TeV J2032+4130.

## 2. Observation & Analysis

TeV J2032+4130 was observed by Chandra on 2004 July 12 at UT 02:04:33 for a total effective exposure time of 48728s using the ACIS-I detector in its standard Timed Exposure Very Faint mode. All four imaging chips were active, in addition to spectroscopy chips S3 and S4. As in the case of our shorter exploratory observation described by Butt et al (2003), the observation was centered on J2000 coordinates RA, Dec=(20:32:07.0, +41:30:30.0). Note that the best-fit location and Gaussian extent of TeV J2032+4130 source have changed slightly between the earlier HEGRA report (Aharonian et al. 2002) and the most recent one (Aharonian et al., 2005) – the current best fit values being: center at RA, Dec=(20:31:57.0s±6.2s stat ±13.7s sys, +41:29:56.8 ± 1.1′stat ± 1.0′sys) with a standard deviation of the 2D Gaussian, $\sigma$=6.2′±1.2′stat±0.9′sys, which are consistent with the previous report.

Pipeline-processed data were prepared using standard techniques and were analyzed using the Chandra Interactive Analysis of Observations (CIAO) software version 3.2.2[3]. The image obtained from the ACIS-I photon event list is illustrated in Figure 1, upon which the TeV source region from Aharonian et al. 2005 has been superimposed. Immediately apparent when comparing this image to that of the shorter exposure presented by Butt et al (2003) is the much larger number of point-like sources that can be seen by eye.

---

[3] For information on CIAO please see http://cxc.harvard.edu/ciao/

The ACIS-I data were further processed using the XPIPE software constructed from CIAO tools and described in detail by Kim et al. (2004). This software produces, among other products, source lists based on the wavelet source detection algorithm WAVDETECT. A list of detected sources, together with their observed counts, are available in tabular form on-line[4].

In order to investigate the hardness of the diffuse emission (including emission from unresolved weaker point sources) within the TeV source region, we used the CIAO tool DMFILTH. This tool uses source and background regions to remove detected source counts and to interpolate over the source region based on information contained within background regions. For reasons of computational expediency, we employed a binning by 4 pixels in both image dimensions for these calculations. This process was carried out for images made from event lists filtered in energy ranges 0.5-2 keV and 2-10 keV. These particular energy ranges were chosen so as to contain approximately half of the X-ray photon events in each. The resulting images were then binned by a further factor of 32 in order to accumulate sufficient counts in each bin to estimate a meaningful hardness ratio. A hardness ratio image was obtained from the ratio of the resulting binned images and is illustrated, together with the TeV source error circle, in Figure 2.

---

[4] Insert sourcelist "Table 0" URL here

The total X-ray flux within the TeV source error circle was estimated based on a weighted effective area computed using the CIAO ACISSPEC utility. The weighted area accounts for spatial quantum efficiency variations and vignetting effects over diffuse source regions. In order to make an accurate flux assessment it is also essential to take into account background events. The ACIS background consists of a relatively soft cosmic X-ray background contribution together with cosmic ray-induced events with a hard spectrum (see, eg., Markevitch et al., 2003). This combined background was estimated for our observation using the Chandra calibration database background samples obtained from composite blank sky fields from which point sources have been removed. The total (diffuse + point-like) background-subtracted TeV source region X-ray flux obtained was $2.9 \times 10^{-12}$ erg cm$^{-2}$ sec$^{-1}$ in the 0.5-5 keV band and $1.4 \times 10^{-12}$ erg cm$^{-2}$ sec$^{-1}$ in the energy range 0.5-2.5 keV, both with an approximate uncertainty of order 10%. At energies above 5 keV the spectrum begins to be background-dominated.

## 3. Cross-correlation with Stellar, Radio & Infrared (2MASS) sources

We cross-correlated our 240 detected X-ray sources with known stars in the field to search for coincidences. We considered the stellar lists included in the Simbad[5] database and found 36 coincidences (Table 1), twelve of which are known OB stars – roughly half as many X-ray/OB stellar coincidences as Albacete Colombo et al. (2005) find in the same-sized field in the core of Cyg OB2. As in the latter study, it is likely that the many of the unidentified X-ray sources in our field are also stars and some may also be X-ray binaries or background extragalactic sources.

---

[5] The Simbad (Set of Identifications, Measurements and Bibliography for Astronomical Data) database can be found on the world wide web at: http://simbad.u-strasbg.fr/Simbad

Setia Gunawan et al. (2003) recently carried out a comprehensive radio survey of the Cygnus OB2 association. We cross-correlated their 350 MHz and 1.4 GHz radio sources (see their Table 2) with the 240 X-ray point sources we detected. The extension of the radio sources were considered in the following way: the observing radio beam size, $\theta_{beam}$, is quoted as 13" at 1.4 GHz and 48" at 350 MHz. Setia Gunawan et al. characterize each source with a flag that relates the source size with the radio beam: flag = 1, 2, 3, 4, and 5 corresponds to radio source sizes less than 1, 1.3, 1.6, and 1.9$\theta_{beam}$, respectively, plus extended sources (ES) of size larger than 1.9$\theta_{beam}$. We took the maximum sized error box consistent with the above flags in checking for radio-X-ray coincidences and a size of 3.0$\theta_{beam}$ for ES sources. The errors in the X-ray sources positions were disregarded, because they are on the order of ~1 arcsec.

We find two of our detected X-ray sources lie within the region of two extended radio sources of Setia Gunawan et al., both of which emit in the 1.42GHz and 350MHz bands. However, the X-ray/radio positional-coincidence is strictly valid for only the broader 350MHz sources (Table 2; see also Fig. 11 in Setia Gunawan et al.).

We also checked for coincidences between the 240 Chandra X-ray sources and the near-infrared (NIR; 1.25 $\mu$m- 2.17$\mu$m) two Micron All Sky Survey[6] (2MASS) sources detected over the same area. For this process we considered a given pair of X-ray and 2MASS sources to be associated if they were within 3 arcsec of each other. This coincidence criterion was chosen as a compromise between Chandra's astrometry for on-axis *vs.* off-axis (psf-distorted) sources, and also because X-ray and NIR emitting regions associated with a given object need not be precisely coincident. Out of the 240 X-ray sources we found 130 to be associated with 2MASS sources and have listed them in Table 3, available on-line[7]. This coincidence fraction, (130/240)~0.54, is somewhat lower than that Albacete Colombo et al. (2005) find for their Chandra observation of the core region of Cyg OB2, (692/1003)~0.69. However, this difference could perhaps be explained by their twice-deeper exposure resulting in more X-ray detections, and consequently a higher coincidence fraction, or it may simply reflect the physically distinct populations in the two fields.

---

[6] The 2MASS archive is accessible on the world wide web via http://irsa.ipac.caltech.edu/

**4. Discussion**

We find no compelling diffuse X-ray counterpart of the extended source, TeV J2032+4130, in a 50 ksec CHANDRA exposure. The total (diffuse + point-like) background-subtracted X-ray flux of the TeV source region is $2.9 \times 10^{-12}$ erg cm$^{-2}$ sec$^{-1}$ in the 0.5-5 keV band and $1.4 \times 10^{-12}$ erg cm$^{-2}$ sec$^{-1}$ in the energy range 0.5-2.5 keV. The hardness image (Fig. 2) reveals no significant hardness increase in the diffuse X-ray flux from the TeV source region. Such hard diffuse emission may be expected if, for instance, the TeV flux were generated by a population of relativistic electrons accelerated in a supernova remnant (SNR) shock. (However, see Chu 1997 for a detailed discussion of the variety of SNR signatures possible in OB associations).

---

[7] Insert Table 3 URL here

As we argued in our previous paper (Butt et al. 2003), we suspect that the TeV emission is related to the young, massive and powerful Cyg OB2 association, and especially to the outlying sub-group of massive stars shown in Fig 1 of that report.

It is possible that this TeV source is a composite made up of many smaller TeV sources that only masquerade as an extended source due to the point spread function (psf) of the HEGRA array (~3 arcmin at 1 TeV; Aharonian et al., 2004). In this case, some subset of the 240 X-ray sources we detect could be viable counterparts to close-by TeV sub-sources that collectively make up TeV J2032+4130. In fact, the surface density of the point-like X-ray sources we detect does reflect an excess consistent with the size and position of the extended TeV source (Figure 3). It is interesting that the radial profile of TeV J2032+4130 (Fig. 1 in Aharonian et al., 2005) also seems to indicate a fairly flat TeV emissivity out to about 7 arcmin from the center of gravity of the source and not a particularly centrally-peaked source (although, admittedly, the large error bars do not permit a definitive view on the detailed morphology of the TeV emission).

However, it cannot be that TeV J2032+4130 is *only* related to the observed surface density of point-like X-ray sources in this region since Albacete Colombo et al. (2005) have observed the core region of Cygnus OB2 only ~20 arcmin distant with Chandra and report 1003 point-like X-ray sources in the same-sized (ACIS-I) field where we find 'just' 240. (Their twice deeper exposure of ~100ksec cannot alone explain the large excess of point-like X-ray sources they report in their Chandra field). There must be something else special about the location of the TeV source region besides the high surface density of X-ray point sources, since it is even higher very nearby without that region being a (yet detected) TeV source.

In fact, Domingo-Santamaria & Torres (2005) (see also, Torres, Domingo-Santamaria & Romero, 2004, Reimer, 2003 and Reimer, Pohl & Reimer, 2005) have very recently published a study showing how cosmic-ray (CR) modulation by powerful stellar winds could lead to a scenario where one would expect TeV-range gamma-ray emission from hadronic target regions located about 1pc from hot OB stars, with little or no MeV-GeV range gammas co-produced[8]. At a distance of 1 pc they showed that the attenuation of TeV range gamma-rays (due to pair-production interactions on the stellar photon bath) was low enough such that they would mostly escape and be visible to us on earth. However, if the target star was in the core region of a dense association, then the intense collective stellar photon fields there would render that region opaque to TeV range gamma-rays (see Reimer 2003 for a more detailed discussion).

Thus, if the volumetric density of OB stars falls off less rapidly with radial distance from the association core, than the volumetric density of all stars, we hypothesize that there ought to be a critical distance from the cluster center where the TeV production via the CRs impinging on target regions ~1pc from hot OB stars outweighs the TeV opacity due to combined cluster stellar photon bath (Fig. 4). (Our hypothesis assumes that the

---

[8] In fact, even if MeV-GeV gammas were produced by this scenario it is not problematic in the case of Cyg OB2: indeed, such a mechanism may well be contributing to the adjacent EGRET source 3EG J2033+4118, positionally coincident with the core of Cyg OB2 association. The extrapolated EGRET range flux from the TeV source region is about a hundredth that of 3EG J2033+4118 so it would not be especially discernable at EGRET sensitivities. Ascertaining whether lower energy GeV range gammas are also emanating from the TeV source region must await the sensitivity and spatial resolution of GLAST.

*intrinsic* TeV emissivity scales with the volumetric density of OB stars – where the TeV gamma-rays are made – and that the TeV opacity scales with the total volumetric density of all stars – due to their combined stellar photon baths.) This differential rate of decrease of the OB *vs.* total stellar population with radial distance is indeed the case for Cyg OB2 – see Fig 6 of Knodlseder 2000. However, at that critical distance from the cluster center there must be a sufficient concentration of hot OB stars present to create a detectable TeV flux, and, as shown in Fig. 1 of Butt et al. (2003) there is indeed a local overdensity of (cataloged) OB stars coinciding with TeV J2032+4130, ~20 arcmin (~10pc) from the cluster core. Note that in other stellar associations the mass segregation may be reversed: more massive stars may cluster more strongly towards the center as compared with their less massive counterparts (see, eg. Fig 6 in Bonnell & Davies, 1998).

Of course, it is also entirely possible that TeV J2032+4130 is unrelated to Cygnus OB2 and that the overdensity of X-ray point sources in the TeV source region is simply a chance effect. But no matter what the ultimate origin of the TeV radiation, if the low measured X-ray and radio emissivity reflects the true intrinsic emission, we favor a nucleonic rather than electronic origin of the very high energy gamma-ray flux (see section 4 of Butt et al. 2003 for further details). On the other hand, it is possible that the intrinsic X-ray emission from the source may be significantly attenuated due to the large amount of interstellar gas and dust in this direction, especially if the TeV source is actually located in the Perseus spiral arm far beyond Cygnus OB2 (see, eg., Fig 10 of Molnar et al., 1995). From the NRAO survey we obtain a value of approximately $1.5 \times 10^{22}$ cm$^{-2}$ for the neutral hydrogen column density in the direction of the TeV source

(Dickey & Lockman, 1990). We caution that this number is obtained by interpolating four neighboring 1x1 degree bins and is thus quite rough. In fact, Scappini et al. (2002) and Casu et al. (2005) have discovered clumpy clouds along lines-of-sight close to the TeV source[9] (eg. towards Cyg OB2 no. 5 at $\alpha=20^h\ 32^m\ 22^s$; $\delta=41^o\ 18'\ 19"$). If the true column density in the direction of the TeV source were higher than the interpolated average of $1.5 \times 10^{22}$ cm$^{-2}$, say, by a factor of ~100 then the intrinsic X-ray emission (0.2-5 keV) may be attenuated by a factor of ~1000 (assuming a power-law X-ray spectrum of index= -2), and a leptonic origin of the TeV radiation could not be ruled out. Of course, this is unlikely to be the case if the TeV source is indeed associated with Cyg OB2 since we do detect many of the stars there in X-rays. However, the TeV source may well be situated much further away, eg. at the distance of Cyg X-3 at ~10kpc, and intervening dense clouds beyond Cyg OB2 cannot be ruled out.

## 5. Conclusions

In summary, we conclude that TeV J2032+4130 is:

- Probably related to some subset of the multiple stellar X-ray sources associated with Cyg OB2 that are clustered in the central region of our Chandra field, consistent with the position and extent of TeV J2032+4130 (Fig. 3 & 4; see also Fig 1 from Butt et al., 2003).

---

[9] These clouds could also serve as targets for high-energy cosmic rays accelerated in Cyg OB2 instead of, or in addition to, the dense wind regions described by Domingo-Santamaria & Torres (2005) and Torres, Domingo-Santamaria & Romero (2004).

- steadily emitting in TeV gamma-rays on years' timescale (Aharonian et al., 2002, 2005), but may perhaps suffer outbursts on decadal timescales which could increase its intensity (eg. Neshpor et al., 1995; see also Lang et al., 2004).
- possibly hadronic in origin, but may be electronic (or even both), depending on the exact amount of X-ray attenuation along the line-of-sight, and the actual distance to the source.

About three shorter mosaic Chandra pointings of fields immediately adjacent to ours would be necessary (and sufficient) to confirm whether the excess X-ray point source surface density in the region truly correlates with the TeV source (Fig. 3). This would be invaluable in making – or breaking – a case for a physical connection between the two. Deeper infrared surveys of this region would also be very helpful in producing a more complete stellar census of the massive stars in Cygnus OB2. Only about 110 of the presumed ~2600 OB stars (Knodlseder 2003) in this stellar association are as yet catalogued.

*Note added in proof*: Albacete Colombo et al. will shortly publish a full report on their Chandra observation of the core region of Cyg OB2 which ought to be very interesting. Also, our colleague Olaf Reimer and co-workers are independently analyzing the Chandra data presented here – we look forward to reading their results. Stay tuned.

**Acknowledgements**

We thank the referee for quick and constructive comments on the original manuscript, which considerably improved the final version. We are grateful to Ettore Flaccomio,

Margaret Hanson and Jürgen Knödlseder for insightful discussions and useful information and figures regarding Cygnus OB2. We thank our colleagues Fernando Comeron, Michael Corcoran, Peter Milne, Martin Pohl, Olaf Reimer, Michael Rupen, and Diego Torres who all participated in the Chandra proposal. YMB is supported by NASA/Chandra and NASA/INTEGRAL General Observer Grants and a NASA Long Term Space Astrophysics Grant. FM acknowledges support by the Swiss Institute of Technology through a Zwicky Prize Fellowship. GER is supported by CONICET (PIP 5375) and ANPCyT (PICT 03-13291), Argentina. This publication makes use of data products from the Two Micron All Sky Survey, which is a joint project of the University of Massachusetts and the Infrared Processing and Analysis Center/California Institute of Technology, funded by the National Aeronautics and Space Administration and the National Science Foundation. We also used the SIMBAD database, operated by CDS, at Strasbourg, France.

**Table 1.** Stellar counterparts to Chandra sources

| X Id. | RA, Dec (J2000) | src | HR | Counterpart | RA, Dec (J2000) | Sp.Type |
|---|---|---|---|---|---|---|
| 2 | 20 32 45.598, 41 25 38.28 | 46 | -1.292 | MT 317, RLP 29 | 20 32 45.44, 41 25 37.5 | O8 V: |
|   |   |   |   | RLP 515, BD+40 4221 | 20 32 45.4, 41 25 37 |   |
| 4 | 20 31 49.678, 41 28 26.04 | 20 | -1.086 | MT 145, RLP 560 | 20 31 49.74, 41 28 26.9 | O9.5V |
| 5 | 20 32 27.842, 41 28 51.95 | 28 | -1.059 | MT 259, RLP 502 | 20 32 27.85, 41 28 52.0 | B0.5V |
| 6 | 20 32 14.885, 41 33 20.89 | 20 | -1.055 | RLP 229 | 20 32 15.10, 41 33 21.0 |   |
| 9 | 20 32 46.319, 41 36 16.20 | 140 | -1.011 | MT 321, BD+41 3798 | 20 32 46.24, 41 36 16.0 |   |
| 11 | 20 32 06.240, 41 24 34.56 | 52 | -0.989 | MT 197, RLP 547 | 20 32 06.19, 41 24 35.8 |   |
|   |   |   |   | MT 198 | 20 32 06.17, 41 24 39.3 |   |
| 12 | 20 32 16.560, 41 25 35.77 | 43 | -0.987 | EM* CDS 1172 | 20 32 16.60, 41 25 36.0 | B(Hα) |
|   |   |   |   | RLP 14 | 20 32 16.50, 41 25 36.0 | O... |
|   |   |   |   | MT 227, RLP 31 | 20 32 16.62, 41 25 36.4 | O9 V |
| 15 | 20 32 13.923, 41 27 11.88 | 233 | -0.968 | CYG OB2 No. 5 | 20 32 13.82, 41 27 11.99 | O7 Ianfp |
| 17 | 20 32 25.438, 41 34 01.56 | 63 | -0.964 | MT 249, RLP 231 | 20 32 25.40, 41 34 02.0 |   |
| 18 | 20 32 27.594, 41 26 21.84 | 120 | -0.961 | MT 258, RLP 30 | 20 32 27.66, 41 26 22.1 | O8 V |
| 19 | 20 31 37.202, 41 23 36.23 | 122 | -0.958 | MT 115, RLP 581 | 20 31 37.40, 41 23 35.5 |   |
| 24 | 20 31 51.360, 41 23 23.28 | 631 | -0.911 | MT 152, RLP 551 | 20 31 51.40, 41 23 23 |   |
| 25 | 20 32 19.925, 41 33 55.08 | 14 | -0.904 | RLP 230 | 20 32 19.90, 41 33 55 |   |
| 26 | 20 32 38.643, 41 25 13.44 | 164 | -0.887 | MT 299, TYC 3161-1029 | 20 32 38.58, 41 25 13.7 | O7.5 V |
| 27 | 20 31 53.757, 41 37 28.92 | 48 | -0.883 | MT 161, RLP 197 | 20 31 53.81, 41 37 29.3 |   |
| 30 | 20 32 14.397, 41 26 34.80 | 20 | -0.846 | MT 218, RLP 526 | 20 32 14.22, 41 26 32.0 |   |
| 33 | 20 32 27.361, 41 29 40.92 | 21 | -0.811 | MT 257, RLP 499 | 20 32 27.33, 41 29 41.2 |   |
| 42 | 20 32 38.876, 41 25 20.65 | 159 | -0.752 | MT 300 | 20 32 38.96, 41 25 20.8 | B |
| 47 | 20 32 13.443, 41 27 11.52 | 118 | -0.725 | Cyg OB2 No. 5 | 20 32 13.823, 41 27 11.99 | O7 Ianfp |
| 53 | 20 32 21.360, 41 28 25.33 | 43 | -0.704 | RLP 531 | 20 32 21.40, 41 28 26 |   |
| 78 | 20 32 38.402, 41 28 56.99 | 52 | -0.631 | MT 298, RLP 495 | 20 32 38.10, 41 28 57 |   |
| 84 | 20 32 37.915, 41 28 52.68 | 105 | -0.597 | MT 298, RLP 495 | 20 32 38.10, 41 28 57 |   |
| 86 | 20 31 50.880, 41 31 22.08 | 32 | -0.592 | RLP 208 | 20 31 50.9, 41 31 18 |   |
| 119 | 20 31 51.360, 41 29 52.43 | 25 | -0.474 | MT 153, RLP 557 | 20 31 51.43, 41 29 52.9 |   |
| 129 | 20 32 13.195, 41 27 24.12 | 82 | -0.427 | MT 213 | 20 32 12.8, 41 27 26 | B0 Vp |
| 141 | 20 32 29.043, 41 25 48.36 | 85 | -0.384 | RLP 512 | 20 32 29.0, 41 25 50 |   |
| 148 | 20 32 10.078, 41 30 18.72 | 40 | -0.371 | MT 207, RLP 535 | 20 32 10.03, 41 30 19.1 |   |
| 162 | 20 31 33.604, 41 28 56.99 | 20 | -0.315 | MT 105, RLP 570 | 20 31 33.66, 41 28 57.5 |   |
| 172 | 20 31 27.595, 41 29 17.52 | 40 | -0.274 | MT 94, RLP 572 | 20 31 27.86, 41 29 16.9 |   |
| 186 | 20 31 37.442, 41 24 15.48 | 39 | -0.203 | MT 116 | 20 31 37.42, 41 24 13.4 |   |
| 199 | 20 31 54.244, 41 34 39.00 | 15 | -0.140 | RLP 202 | 20 31 54.2, 41 34 40 |   |
| 210 | 20 31 35.039, 41 29 32.27 | 30 | 0.101 | MT 111, RLP 569 | 20 31 34.81, 41 29 33.0 |   |

**Table 2.** Chandra X-ray/Radio coincidences

| X-ray Id. | X-ray position RA, Dec (J2000) (hms,dms) | Radio Id. | Radio position RA, Dec (J2000) (hms,dms) | Radio flux 350 MHz (mJy) |
|---|---|---|---|---|
| 36 | 20 32 3.122, 41 38 23.64 | 218 | 20 32 2.16, 41 37 59.24 | 122±3 |
| 153 | 20 31 58.32, 41 36 48.60 | 217 | 20 32 1.22, 41 37 13.64 | 87±5 |

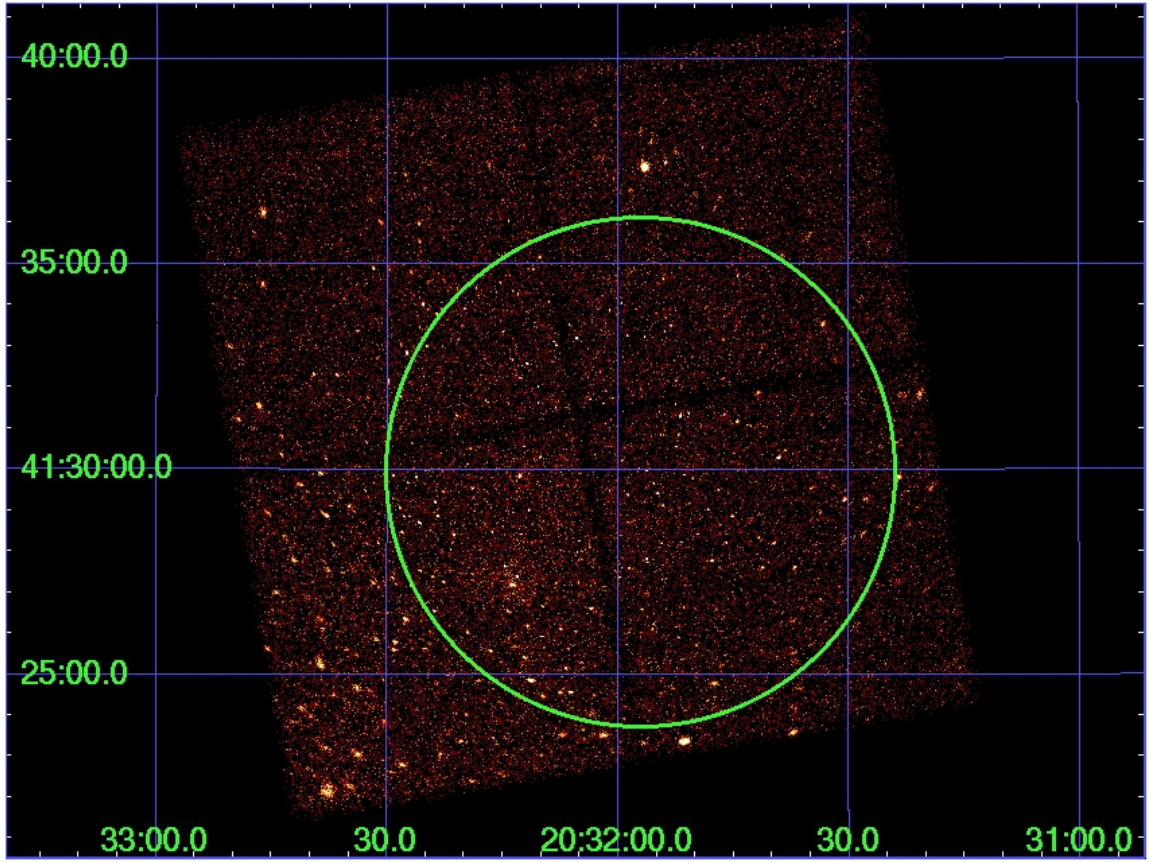

Fig 1: 50 ksec CHANDRA image of the 4 ACIS I-array chips. The circle shows the ~6' (1$\sigma$) extent of the extended TeV source, TeV J2032+4130, reported by the HEGRA collaboration (Aharonian et al., 2005). The aim point is at $\alpha$(J2000)=$20^h32^m07^s$, $\delta$(J2000)=+41°30'30". North is up; east is to the left. The best-fit location and Gaussian extent of the TeV source changed slightly between the earlier HEGRA report and the most recent one (Aharonian et al., 2002 *vs.* 2005) which is why the superimposed circle appears slightly off-center (see text).

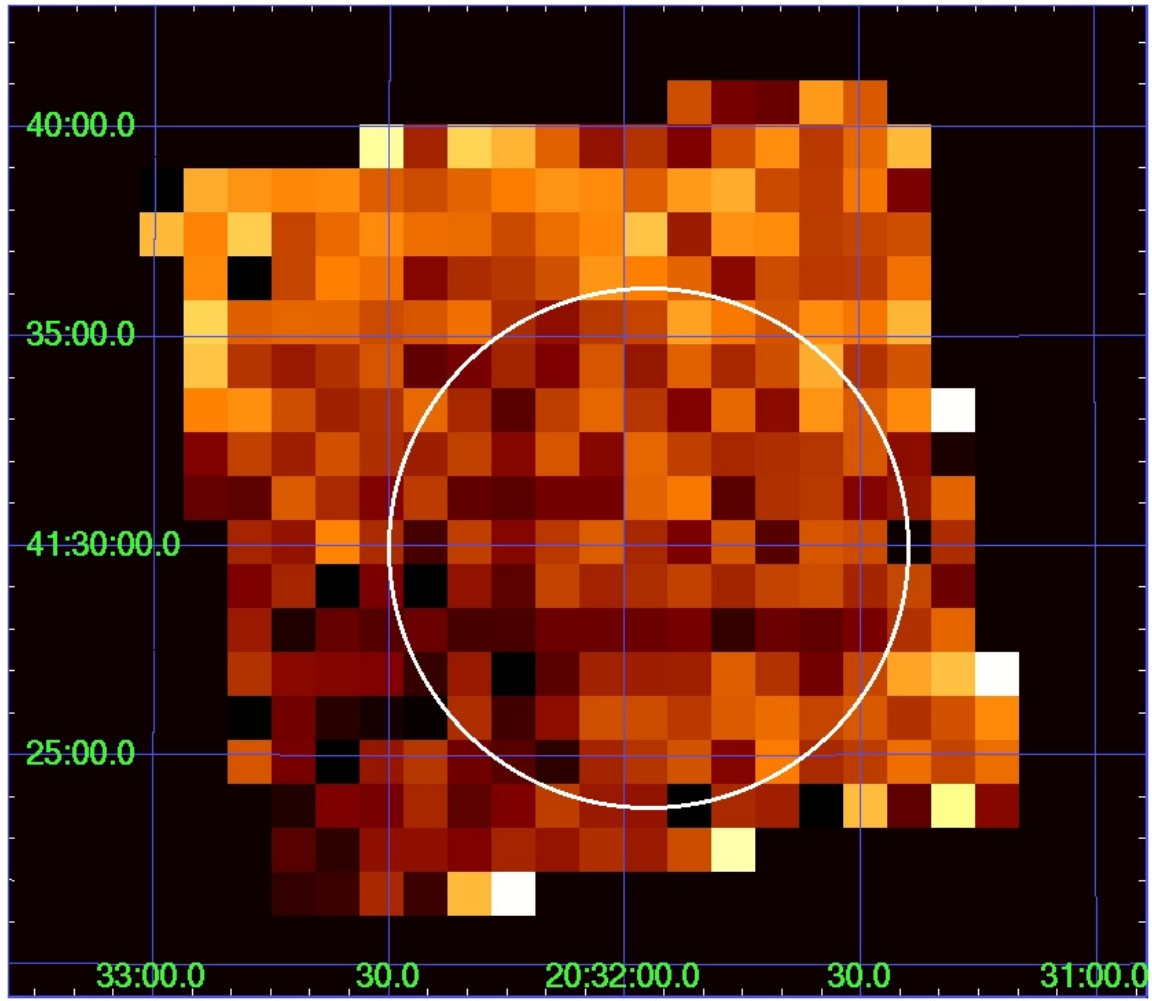

Fig 2: A hardness image (2-10 keV)/(0.5-2keV) of the same field of view as in Fig 1 – Brighter pixels are harder. The image has been binned by 32 pixels in order to accumulate sufficient counts in each bin to estimate a meaningful hardness ratio. The circle shows the ~6' (1$\sigma$) extent of the extended TeV source, TeV J2032+4130, reported by the HEGRA collaboration (Aharonian et al., 2005). The aim point is at $\alpha$(J2000)=20$^h$32$^m$07$^s$, $\delta$(J2000)=+41$^o$30'30". North is up; east is to the left. The best-fit location and Gaussian extent of the TeV source changed slightly between the earlier HEGRA report and the most recent one (Aharonian et al., 2002 *vs.* 2005) which is why the superimposed circle appears slightly off-center (see text).

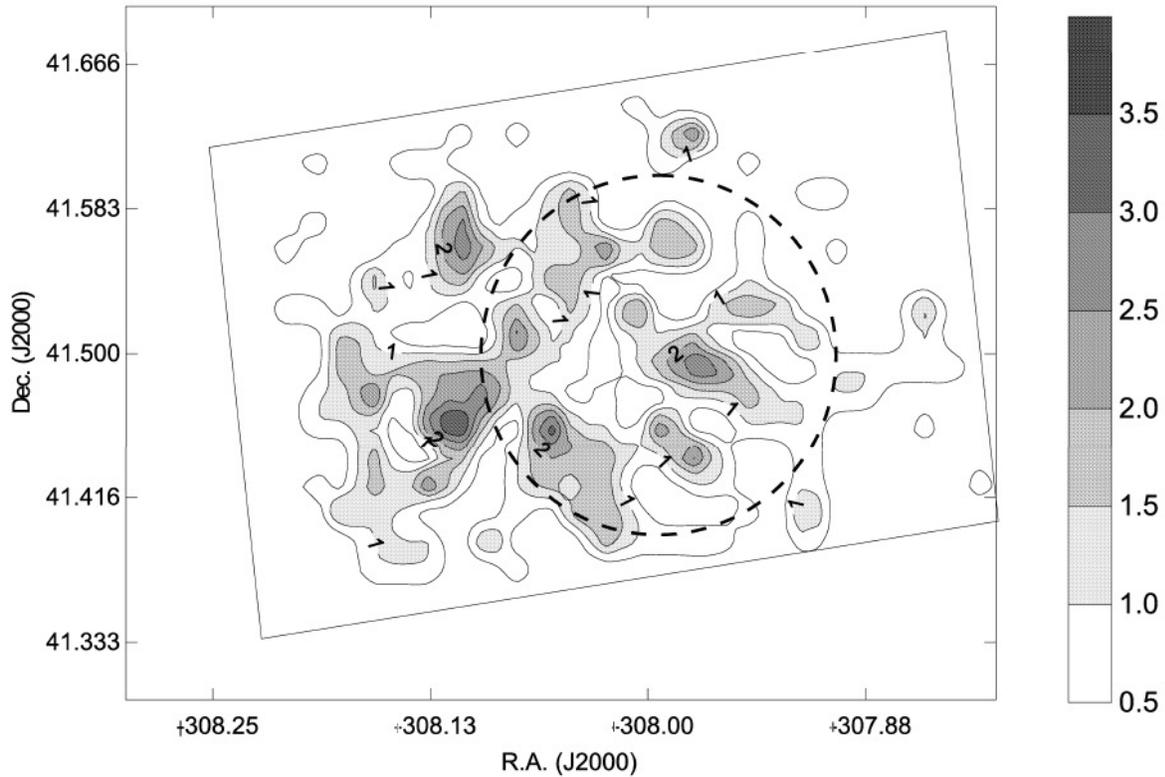

Fig 3: A surface density plot of the 240 point-like X-ray sources detected in our Chandra Observation of TeV J2032+4130. The cell-size used for the smoothing was 1′ x 1′. The greyscale bar at right indicates the number of point-like X-ray sources per arcmin$^2$. The dotted circle shows the 1$\sigma$ extent of TeV J2032+4130 and the slanted inner rectangular region outlines the Chandra ACIS-I field of view. The X-ray point-source distribution is far from uniform: distinct concentrations of sources within the central ~7′ of the field of view are revealed. Taken together, these local maxima are consistent with the size and location of TeV J2032+4130, given the generous uncertainties quoted in its location and extent: RA, Dec=(20:31:57.0s±6.2s stat ±13.7s sys, +41:29:56.8 ± 1.1′ stat ± 1.0′ sys) with a standard deviation of the 2-dimensional Gaussian, $\sigma$=6.2′ ±1.2′ stat±0.9′ sys (Aharonian et al., 2005).

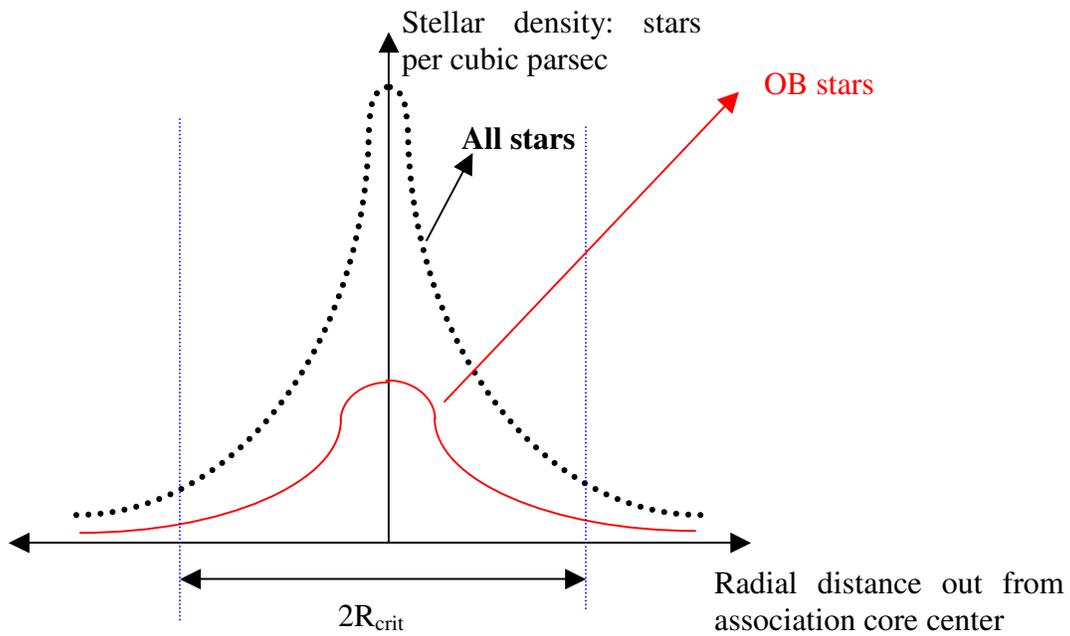

Fig 4: A hypothesis for the origin of TeV J2032+4130 based on the work of Reimer (2003) and Domingo-Santamaria & Torres (2005). Plotted are the stellar volumetric densities for all stars (dotted) and that for OB stars alone (solid). The collective stellar photon bath quenches any intrinsic TeV emission produced at $r<R_{crit}$ due to pair-production. However, due to the more rapid decrease of the all-stellar *vs.* OB-only volumetric densities with distance from the cluster center, outside of $R_{crit}$, TeV gamma-rays may survive. This figure should be compared to Figure 6 of Knodlseder (2000). In the case of TeV J2032+4030, the TeV emissivity may be especially enhanced due to the outlying concentration of OB stars (Fig 1 of Butt et al., 2003).